\documentclass[a4paper,11pt]{article}
\usepackage{jcappub} 
\usepackage{graphicx}
\usepackage{dcolumn}
\usepackage{bm}
\usepackage[export]{adjustbox}
\everymath{\displaystyle}

\title{Cross terms and monochromatic gravitational wave sources in our Galactic Centre}

\author[a,b,c]{Pau Amaro Seoane,}
\affiliation[a]{Universitat Politècnica de València, Spain}
\affiliation[b]{Max Planck Institute für Extraterrestriche Physik, Garching, Germany}
\affiliation[c]{Higgs Centre for Theoretical Physics, Edinburgh, UK} 

\author[a]{Josep V. Arnau and}

\author[a]{Màrius Josep Fullana i Alfonso}

\emailAdd{amaro@riseup.net}

\abstract{
The gravitational capture of a small compact object by a supermassive black hole is one of the most intriguing sources of gravitational waves to be detected by space-borne observatories.
Modeling gravitational waves is a challenging task. However, various approximations exist that enable us to claim detection and perform parameter extraction with high accuracy on synthetic data.
The first numerical implementation of relativistic corrections for an astrophysical system was conducted in a post-Newtonian (PN) framework, and since then, most comparable programmes have followed a similar approach. Nevertheless, the PN approach has been developed for two particles in a vacuum. Here, we present the first results on the impact of cross-terms in the PN scheme for ``monochromatic'' sources (meaning that the peak frequency does not evolve over the observational time), which account for the interplay between different bodies.
We present a conceptual and illustrative study of the cross-terms on the first PN term in a group of three bodies representing asymmetric binaries (X-MRIs, E-EMRIs, and X-MRIs with E-EMRIs). We find that these cross-terms can lead to a complete phase shift of the gravitational wave by a few times $2\pi$ over periods of time of less than one year for the close binaries.
Ignoring the cross-terms in the relativistic semi-Keplerian system can significantly complicate parameter extraction for monochromatic sources.
}

\begin{document}
\maketitle
\flushbottom

\section{Introduction}
\label{sec.intro}

The first implementation of a post-Newtonian correction in a purely Newtonian $N-$body code by \cite{KupiEtAl06} was conducted approximately 17 years ago. In their work, the authors modified the program to incorporate the 1PN, 2PN, and 2.5PN terms to account for the periapsis shift and the loss of energy via gravitational radiation. Later, \cite{BremAmaro-SeoaneSpurzem2013} expanded this work by adding non-spinning post-Newtonian corrections to the accelerations up to 3.5PN. They also included spin-orbit coupling up to next-to-lowest order and the lowest-order spin-spin coupling to integrate spin precession. An alternative to this scheme for the EMRI problem, as seen in the approach followed by \cite{BremAmaroSeoaneSopuerta2014}, involves combining the Newtonian integrator with a geodesic solver when the particle of interest, the EMRI, is decoupled from the stellar system.

Moreover, most current relativistic dynamical approaches rely on the post-Newtonian approximation. In this work, we discuss the r{\^o}le of the cross terms, as introduced in the work of \cite{Will2014}, to the first order in the expansion. This means that we do not consider dissipative terms, allowing us to focus in this article on ``monochromatic'' mass ratio inspirals, both extreme- and extremely large; i.e., E-EMRIs \citep{AmaroSeoaneLinTzanavaris2024} and X-MRIs \citep{Amaro-Seoane2019}. These sources have a very long inspiral time, which implies that their frequency peaks can be considered fixed when comparing the duration of the mission (years) with the lifetime of the source in the band (up to a million years). We therefore focus on these two types of sources and analyse the secular effect of the cross terms on their evolution, using the first post-Newtonian relativistic correction without cross-terms as a reference.

The examples we adopt are representative but ignore the influence that a system of a few tens of X-MRIs around the MBH \citep{Amaro-Seoane2019} could have. In that sense, our integrations illustrate the impact of neglecting the cross terms but should not be taken as absolute. The role of this swarm of X-MRIs will be presented elsewhere soon. Considering that excluding the cross terms can lead to a complete phase shift, new strategies and algorithms must be developed to reliably perform parameter extraction for the holy grail of space-borne observatories: asymmetric binaries of extreme- and extremely large mass ratios, which will enable robust tests of general relativity \citep{Amaro-SeoaneEtAl2017,Amaro-SeoaneGairPoundHughesSopuerta2015,Amaro-SeoaneLRR2012,AmaroSeoane2022}.

\section{Algorithm}

Following the work and nomenclature of \cite{Will2014}, we refer to the black hole as the first body in the summations. We represent it in any quantity, such as the mass, with the subscript $m_{\bullet}$. By considering the conservation of linear momentum at the Newtonian level, expressed as $M{\bm v}_{\bullet} + \sum_a m_a {\bm v}_a = 0$, we can eliminate the velocity ${\bm v}_{\bullet}$ from the post-Newtonian terms in the equations of motion. Consequently, the resulting equations of motion assume the form

\begin{align}
{\bm a}_a = & - \frac{GM{\bm x}_{a\,\bullet}}{r_{a\,\bullet}^3}- \sum_b \frac{Gm_b{\bm x}_{ab}}{r_{ab}^3}\nonumber \\
& + \frac{1}{c^2}[{\bm a}_a]_{\rm PN1} + \frac{1}{c^2}[{\bm a}_a]_{\rm X}
\\
& + O\left (\frac{G^2 m_b^2}{c^2 r^3}\right ) \,,
\label{acc1}
\end{align}

\noindent 
In this equation

\begin{subequations}
\begin{align}
[{\bm a}_a]_{\rm PN1} &= \frac{GM{\bm x}_{a\,\bullet}}{r_{a\,\bullet}^3} \left ( 4\frac{GM}{r_{a\,\bullet}} - v_a^2 \right ) + 4\frac{GM}{r_{a\,\bullet}^3} \left ({\bm v}_a \cdot {\bm x}_{a\,\bullet} \right ) {\bm v}_a \,,
\label{acc1PN}
\\
{[{\bm a}_a]}_{\rm X} 
&= 5\frac{G^2 m_aM {\bm x}_{a\,\bullet}}{r_{a\,\bullet}^4}
- \frac{G m_a }{r_{a\,\bullet}^3} \left [ 4 v_a^2 {\bm x}_{a\,\bullet} - 7( {\bm v}_a \cdot {\bm x}_{a\,\bullet}){\bm v}_a \right ]
\nonumber \\
& \quad 
+\sum_b \frac{G^2 m_bM {\bm x}_{a\,\bullet}}{r_{a\,\bullet}^3} \left ( \frac{4}{r_{ab}} + \frac{5}{4r_{b\,\bullet}} + \frac{r_{a\,\bullet}^2}{4r_{b\,\bullet}^3}  -\frac{r_{ab}^2}{4r_{b\,\bullet}^3} \right )
\nonumber \\
& \quad 
+ \sum_b \frac{G^2 m_bM {\bm x}_{ab}}{r_{ab}^3} \left ( \frac{4}{r_{a\,\bullet}} + \frac{5}{4r_{b\,\bullet}} - \frac{r_{a\,\bullet}^2}{4r_{b\,\bullet}^3}  +\frac{r_{ab}^2}{4r_{b\,\bullet}^3} \right )
\nonumber \\
& \quad
- \frac{7}{2} \sum_b \frac{G^2 m_bM {\bm x}_{b\,\bullet}}{r_{b\,\bullet}^3} \left ( \frac{1}{r_{ab}} - \frac{1}{r_{a\,\bullet}} \right )
\nonumber \\
& \quad
- \sum_b  \frac{G m_b }{r_{a\,\bullet}^3}  \left [ 4 ({\bm v}_a \cdot {\bm v}_b ){\bm x}_{a\,\bullet} \right. \nonumber \\
& \left. \quad -3 ({\bm v}_b \cdot  {\bm x}_{a\,\bullet} ) {\bm v}_a - 4 ({\bm v}_a \cdot  {\bm x}_{a\,\bullet} ) {\bm v}_b \right ]
\nonumber \\
& \quad
+\sum_b \frac{G m_b {\bm x}_{ab}}{r_{ab}^3} \left [ v_a^2 - 2 |{\bm v}_{ab}|^2 + \frac{3}{2} \left ( {\bm v}_b \cdot {\bm n}_{ab} \right )^2 \right ] 
\nonumber \\
& \quad
+\sum_b \frac{G m_b }{r_{ab}^3} \left [ {\bm x}_{ab} \cdot (4{\bm v}_a -3{\bm v}_b ) \right ]{\bm v}_{ab} \,,
\label{acc1cross}
\end{align}
\label{acc1combined}
\end{subequations}

\noindent 
where we use, as in the original work, \({\bm v}_{ab} \equiv {\bm v}_a - {\bm v}_b\), and the summation over
\(b\) excludes both body \(a\) and the black hole. Essentially, we are
neglecting post-Newtonian terms that involve solely star-star interactions; in
this reduced form, the equations encompass only two-body interactions. 

Following the same idea, the equations of motion for the MBH is

\begin{equation}
{\bm a}_{\bullet}= - \sum_b \frac{Gm_b{\bm x}_{\bullet b}}{r_{\bullet b}^3}
+ \frac{1}{c^2}[{\bm a}_{\bullet}]_{\rm PN1} + \frac{1}{c^2}[{\bm a}_{\bullet}]_{\rm X} + 
O\left ( \frac{G^2 m_a^3}{Mc^2r^3} \right) \,,
\label{accH}
\end{equation}

\noindent
where

\begin{subequations}
\begin{align}
[{\bm a}_{\bullet}]_{\rm PN1} &= \sum_b \frac{Gm_b{\bm x}_{\bullet b}}{r_{\bullet b}^3} \left [ 5\frac{GM}{r_{\bullet b}} - 2 v_b^2 + \frac{3}{2} ({\bm v}_b \cdot {\bm n}_{\bullet b} )^2 \right ] 
\nonumber \\
& \quad + 3\sum_b \frac{Gm_b}{r_{\bullet b}^3} \left ({\bm v}_b \cdot {\bm x}_{\bullet b} \right ) {\bm v}_b \,,
\label{accHPN}
\\
{[{\bm a}_{\bullet}]}_{\rm X} 
&= 4\sum_b \frac{G^2m_b^2{\bm x}_{\bullet b}}{r_{\bullet b}^4} 
\nonumber \\
& \quad + \sum_{b,c} \frac{G^2m_b m_c {\bm x}_{\bullet b}}{r_{\bullet b}^3} \left ( \frac{4}{r_{\bullet c}} + \frac{5}{4r_{bc}} - \frac{r_{\bullet c}^2}{4r_{bc}^3} + \frac{r_{\bullet b}^2}{4r_{bc}^3} \right )
\nonumber \\
& \quad -\frac{7}{2} \sum_{b,c} \frac{G^2 m_b m_c {\bm x}_{bc}}{r_{bc}^3 r_{\bullet b}}
\nonumber \\
& \quad - \sum_{b,c} \frac{Gm_b m_c}{Mr_{\bullet b}^3}  \Big[ 4 ({\bm v}_b \cdot {\bm v}_c ){\bm x}_{\bullet b} -3 ({\bm v}_b \cdot  {\bm x}_{\bullet b} ) {\bm v}_c  \nonumber \\  
& \quad  - 4 ({\bm v}_c \cdot  {\bm x}_{\bullet b} ) {\bm v}_b \Big] \,.
\label{accHcross}
\end{align} 
\label{accHcombined}
\end{subequations}

\noindent 
These equations of motion can be derived from the truncated Lagrangian and Hamiltonian, as provided by the work of \cite{Will2014}. In turn, they allow for the existence of conserved quantities, namely the total energy and momentum, which can be expressed to the relevant order as follows:

\begin{subequations}
\begin{align}
&E = \frac{1}{2} Mv_{\bullet}^2 + \frac{1}{2} \sum_a m_a v_a^2 
- \frac{1}{2} \sum_{a,b} \frac{Gm_a m_b}{r_{ab}} - \sum_a \frac{GMm_a}{r_{\bullet a}}
\nonumber \\
& \quad + \frac{1}{c^2} \biggl \{ \frac{3}{8} \sum_a m_a v_a^4 + \frac{3}{2} \sum_a \frac{GMm_a}{r_{\bullet a}} v_a^2 
+ \frac{1}{2} \sum_a \frac{G^2 M^2 m_a}{r_{\bullet a}^2}
\nonumber \\
& \quad \quad
+ \frac{1}{4} \sum_{a,b} \frac{G m_a m_b}{r_{ab}} \left [ 6v_{a}^2 - 7 {\bm v}_a \cdot {\bm v}_b - ({\bm n}_{ab} \cdot {\bm v}_a )({\bm n}_{ab} \cdot {\bm v}_b ) \right ]
\nonumber \\
& \quad \quad
+ \frac{1}{2} \sum_{a} \frac{G M m_a }{r_{\bullet a}} \left [ 3v_{\bullet}^2 - 7 {\bm v}_{\bullet}\cdot {\bm v}_a - ({\bm n}_{\bullet a} \cdot {\bm v}_{\bullet})({\bm n}_{\bullet a} \cdot {\bm v}_a ) \right ]
\nonumber \\
& \quad \quad
+ \sum_{a,b} \frac{G^2 Mm_a m_b }{r_{ab} r_{\bullet a}}
+ \frac{1}{2} \sum_{a,b} \frac{G^2 Mm_a m_b }{r_{\bullet a} r_{\bullet b}}
\biggr \} + O\left (\frac{G^2 m_a^3}{r^2} \right)\,, 
\label{eq.RelEnergy}
\\
& {\bm P} = M{\bm v}_{\bullet}+ \sum_{a} m_a \bm{v}_a \left ( 1 + \frac{1}{2c^2} v_a^2 \right )
\nonumber \\
& \quad - \frac{1}{2c^2} \biggl \{ \sum_a \frac{GMm_a}{r_{\bullet a}} \left [ {\bm v}_a + ({\bm v}_a \cdot {\bm n}_{\bullet a}) {\bm n}_{\bullet a} \right ]
\nonumber \\
& \quad \quad + \sum_a \frac{GMm_a}{r_{\bullet a}} \left [ {\bm v}_{\bullet}+ ({\bm v}_{\bullet}\cdot {\bm n}_{\bullet a}) {\bm n}_{\bullet a} \right ]
\nonumber \\
& \quad \quad + \sum_{a,b} \frac{Gm_a m_b}{r_{ab}} \left [ {\bm v}_a + ({\bm v}_a \cdot {\bm n}_{ab}) {\bm n}_{ab} \right ] \biggr \}
\nonumber \\
& \quad + O\left (\frac{Gm_a^3 v_a}{c^2 Mr} \right ) \,,
\end{align}
\end{subequations}

\noindent 
and we note that the summations exclude the MBH. 

So as to evaluate the contribution of the cross terms as compared to the post-Newtonian one, we define the following ratios, $R$, 

\begin{align}
[R_{\bullet}]_{\rm PN1} & := \left | \frac{[{\bm a}_{\bullet}]_{\rm PN1}}{- c^2 \sum_b {Gm_b{\bm x}_{\bullet b}}/{r_{\bullet b}^3}} 
\right | \nonumber \\
[R_{\bullet}]_{\rm X} & := \left | \frac{[{\bm a}_{\bullet}]_{\rm X}}{- c^2 \sum_b {Gm_b{\bm x}_{\bullet b}}/{r_{\bullet b}^3}} \right |
\end{align}

\noindent
for the MBH, and 

\begin{align}
[R_a]_{\rm PN1} & := \left | \frac{
[{\bm a}_a]_{\rm PN1}}{ c^2 \left ( - {GM{\bm x}_{a\,\bullet}}/{r_{a\,\bullet}^3}- \sum_b {Gm_b{\bm x}_{ab}}/{r_{ab}^3} \right )} \right | \label{eq.RaPN1} \\
[R_a]_{\rm X} & := \left | \frac{
[{\bm a}_a]_{\rm X}}{ c^2 \left ( - {GM{\bm x}_{a\,\bullet}}/{r_{a\,\bullet}^3}- \sum_b {Gm_b{\bm x}_{ab}}/{r_{ab}^3} \right )} \right | \label{eq.RaX}
\end{align}

\noindent
for the CO $a$. Note that those ratios are the comparison of each first order PN contribution respect to the Newtonian terms (the zero order ones) that acts on each body at each step of the integration. 

\section{Impact on the evolution of asymmetric binaries}

In this section, we illustrate the relevance of the cross terms on the evolution of different systems. We note that these are representative, idealised examples. Again, as explained in the introduction, it is important to emphasise that, in particular, we neglect the effect that the presence of a system of a few tens of X-MRIs around the MBH \citep{Amaro-Seoane2019} could have on the systems we are considering. This swarm of X-MRIs has a significant impact on the system, which will be addressed in detail elsewhere soon.

To perform the calculations, we have adopted the direct-summation $N-$body code \texttt{planet} by Sverre Aarseth, a programme designed to follow the evolution of small bodies (planets) revolving around a massive one (a star) with high accuracy. The program only considers dynamics (i.e., pure point-like gravitational effects) and is particularly suited to handle extreme mass ratios and a very large number of orbits, ensuring that secular effects are efficiently captured in the integration. This is not the first time that \texttt{planet} has been used to address asymmetric binaries. The work of \cite{Amaro-SeoaneBremCuadraArmitage2012} already adapted it to include post-Newtonian terms. In contrast to their approach, our version of \texttt{planet} only includes the first post-Newtonian approximation and its cross terms. We do not include any order beyond the work of \cite{Will2014}; this means that we only consider the first contribution to the periapsis shift and do not include any dissipative terms. We also recalculate the positions of the particles relative to the centre-of-mass of the system at every timestep, because in the original version the calculations were done relative to the origin of coordinates.

\subsection{A system of two early EMRIs}

The first example we address is that of two E-EMRIs, as in the work of \cite{AmaroSeoaneLinTzanavaris2024}; i.e., two extreme-mass ratio inspirals that are still in the early stages of their evolution toward becoming polychromatic sources. Since their evolution in phase space is very slow compared to the mission duration, they are ideal sources for this study, as we can neglect the shrinkage of the semi-major axis due to gravitational radiation and focus solely on the periapsis shift.

\begin{figure*}
          {\includegraphics[width=1.0\textwidth,center]{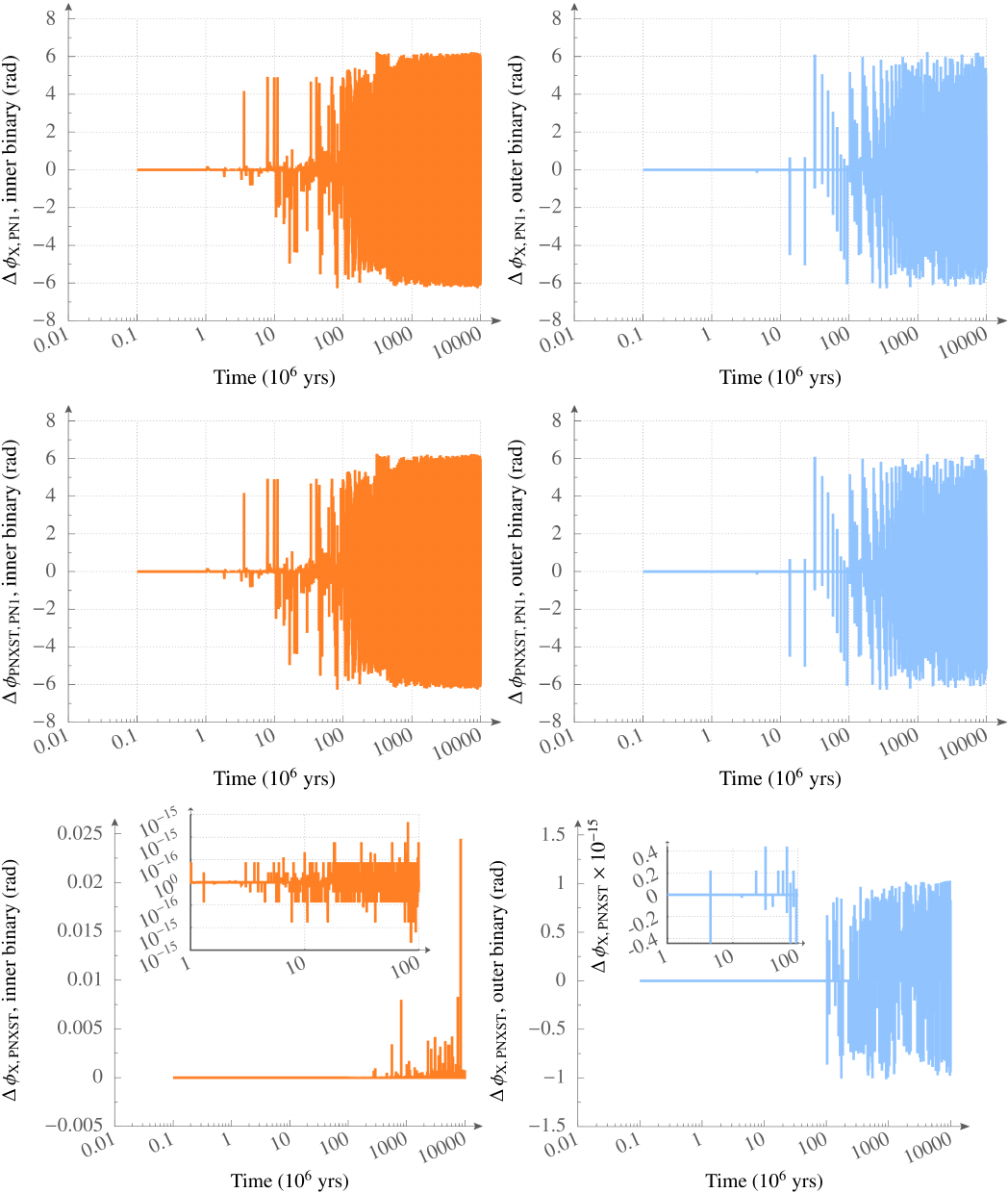}}
\caption
   {
Throughout all of this work, the central supermassive black hole has a fixed mass equivalent to that of SgrA*; i.e., $4.3\times 10^6\,M_{\odot}$. The two E-EMRIs we consider in this figure both have a mass of $10\,M_{\odot}$; the first one has an initial semi-major axis of $10^{-2}\,\text{pc}$ with an eccentricity of $e=0.987$ (i.e., it has evolved toward higher frequencies), while the second one has an initial semi-major axis of $10^{-1}\,\text{pc}$ with an eccentricity of $e=0.9999$. We depict the phase difference in radians as a function of time in years for the inner (left panels, orange) and outer (right panels, blue) binary between the full post-Newtonian correction including the cross terms (subindex ``X'') and the first post-Newtonian term (PN1), upper row. The second row corresponds to the same but taking into account only the stars in the cross terms (``PNXST''), i.e., ${[{\bm a}_a]}_{\rm X}$, Eq.~(\ref{acc1cross}). The last row is the phase difference in the cross terms solution including the MBH and excluding it.
   }
\label{fig.Diff_Phase_E_EMRIs_Aproximats_low}
\end{figure*}

We consider a system of three objects: the central massive black hole (MBH) and two E-EMRIs. Since we have a full Hamiltonian system (masses, positions, velocities), we can derive the phase of the gravitational wave. In figure~(\ref{fig.Diff_Phase_E_EMRIs_Aproximats_low}), we display the evolution in time of the phase difference of the gravitational wave induced on the outer and inner binaries for three different combinations. In the first case, we compare the full cross-terms solution to the usual first post-Newtonian correction, which does not include the interplay of relativistic terms. The second comparison is between the cross-terms solution, ignoring the correction for the central massive black hole (i.e., ${[{\bm a}_a]}_{\rm X}$ as given by Eq.~(\ref{acc1cross})), and again the usual first post-Newtonian correction. Finally, we compare the full cross-terms solution with the approximation in which the contribution from the MBH cross-terms is ignored.

We integrate the system for about two million years because (i) we are interested in the secular, cumulative effect of the phase difference; (ii) these systems, E-EMRIs, remain in the detection band for a very long time, on the order of the integration time in our simulations, as shown by the work of \cite{AmaroSeoaneLinTzanavaris2024}. It is important to note, however, that in that article, the authors did not take into account the interplay with the predicted swarm of X-MRIs encircling the MBH, as predicted by \cite{Amaro-Seoane2019}. Lastly, (iii) the early integration of this system is not representative of what a space-borne observatory such as LISA will observe. Indeed, LISA can observe asymmetric mass ratio inspirals out to a redshift of $\sim 2$ in the most pessimistic case \citep{BabakEtAl2017}. In a volume of that redshift radius, the number of galaxies harbouring an E-EMRI is very large, and statistically, they will have been evolving for a long time. That is, late times in the integration are representative, while early times are not. This is why we focus on typical late integrations.

When calculating the phase shift and other quantities in this article, we have eliminated transient, episodic outliers. For all quantities displayed in the figures, such as in figure~(\ref{fig.Diff_Phase_E_EMRIs_Aproximats_low}), any event at $5\sigma$ in the distribution has been removed, as we consider it statistically spurious and not indicative of the true behavior. Including such events would artificially inflate the results.

We observe that the phase shift for the outer binary is on the order of a few times $2\pi$. This will not only make parameter extraction challenging but may also complicate the detection of the source itself. The mildest difference is between the full cross-terms solution and the full cross-terms solution neglecting the interplay with the MBH. Even in this case, we observe that the phase shift at representative times can become very large. This is significant because one might incorrectly assume that the motion of the MBH compared to the small compact bodies orbiting around it is negligible, and thus the contributing relativistic cross-terms can be ignored. The results of this simulation, and those in the following subsections, demonstrate that this is not the case. The last row of panels shows that the phase shift for the inner binary is very small but not zero. For the outer binary, we see that at later, representative times, the phase shift reaches values of about 1.

\begin{figure*}
          {\includegraphics[width=1.0\textwidth,center]{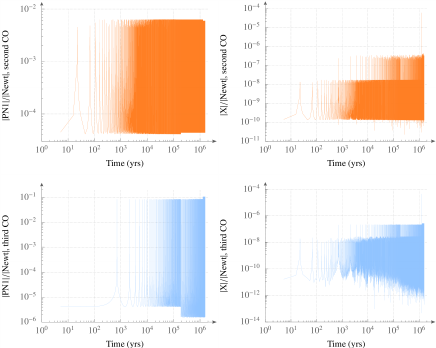}}
\caption
   {
Top left panel: Evolution in time in years of the ratio between the first post-Newtonian correction term and the Newtonian contribution, $[R_a]_{\rm PN1}$, Eq.~(\ref{eq.RaPN1}), for the integration of the orbit of the second compact object (``CO''). Top right panel: Same for the cross terms, $[R_a]_{\rm X}$, Eq.~(\ref{eq.RaX}). Bottom left panel: Same as the top left panel for the third CO. Bottom right panel: Same as the top right panel for the third CO.
   }
\label{fig.Ratios_E_EMRIs_Aproximats_low}
\end{figure*}

In figure~(\ref{fig.Ratios_E_EMRIs_Aproximats_low}), we show the contribution of the post-Newtonian and the full cross-term solution as compared to the Newtonian case. Following the same colour scheme as before, orange denotes the inner object (second CO, since the central MBH is considered to be the first one in the summations) and blue the outer one (third CO). As expected, the contribution of the relativistic terms becomes increasingly important as time passes, since it accumulates in a secular fashion. Additionally, for the farthest object, the ratios are smaller, as the relative velocities decrease.

An important aspect to evaluate in the numerical integration of a dynamical system is the
effect of the numerical error. We use Eq.~(\ref{eq.RelEnergy}) to do so and find that the 
``instantaneous'' energy error (i.e. the error produced between two consecutive timesteps) produced by the corrector algorithm in our simulations is kept at the order of $\sim 10^{-9}$ and below. The predictor algorithm
produces larger energy errors (of the order of $\sim 10^{-6}$) but the code follows the instructions
of the corrector. That means that it does what it has to do, \textit{it corrects} the orbit going
back to the previous timestep and modifies accordingly the orbit. This means that the numerical
error is determined by that of the corrector algorithm. Moreover, even if we guided ourselves wrongly by the predictor error, we stress here that we are not interested in the cumulative error since the beginning of the integration, but in the error in a time span of a few years (the observational time of LISA) when the system has forgotten about its initial conditions. Briefly, the integration error is negligible.

\subsection{One E-EMRI and one X-MRI}

In this subsection, we repeat the same exercise as in the previous one but for a system composed of one E-EMRI and an inner X-MRI. Due to the large number of orbits described by the inner binary, in this case, we had to limit ourselves to relatively short integration times, as the size of the output files became too large to analyse, even when the output was set to large intervals of time.

\begin{figure*}
          {\includegraphics[width=1.0\textwidth,center]{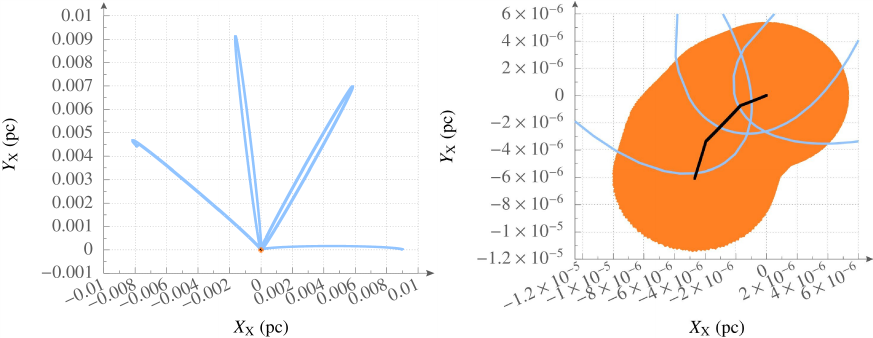}}
\caption
   {
Orbits of the outer (left panel, blue) and inner (right panel, orange) binaries in the X-Y plane, in parsecs, for the full cross-terms relativistic integration (subindex ``X'', cross). The E-EMRI has a mass of $10\,M_{\odot}$, an initial semi-major axis of $4.52\times 10^{-3}\,\text{pc}$, and an initial eccentricity of $e=0.999333$. The X-MRI has a mass of $5\times 10^{-2}\,M_{\odot}$, an initial semi-major axis of $2.88\times 10^{-6}\,\text{pc}$, and an initial eccentricity of $0.84$. The outer binary, the E-EMRI, revolves three times around the MBH, as seen in the left panel, and has a semi-major axis about eight orders of magnitude larger than the inner binary, which is nevertheless visible in the same panel, along with the MBH in black. On the right panel, we can clearly see the orbit of the inner binary, as well as the periapsis passages of the outer binary and the full trajectory of the MBH, which is drifting towards higher values in both X and Y. This is due to the gravitational attraction of the outer binary, which also deviates the trajectory of the inner binary.
   }
\label{fig.Orbits_E_EMRI_X_MRI_X_low}
\end{figure*}

In figure~(\ref{fig.Orbits_E_EMRI_X_MRI_X_low}), we show the orbits in the X-Y plane of the inner and outer binaries. The inner binary corresponds to the X-MRI. We can clearly see how the E-EMRI precesses around the SMBH and completes three periapsis passages. The scale is such that we have added an additional panel to show the evolution of the X-MRI, the inner, orange binary, on the right of the same figure. The X-MRI goes through a much larger number of cycles around the SMBH, which, as in the case of the X-MRI, is drifting towards the right due to the gravitational attraction of the E-EMRI.

\begin{figure*}
          {\includegraphics[width=1.0\textwidth,center]{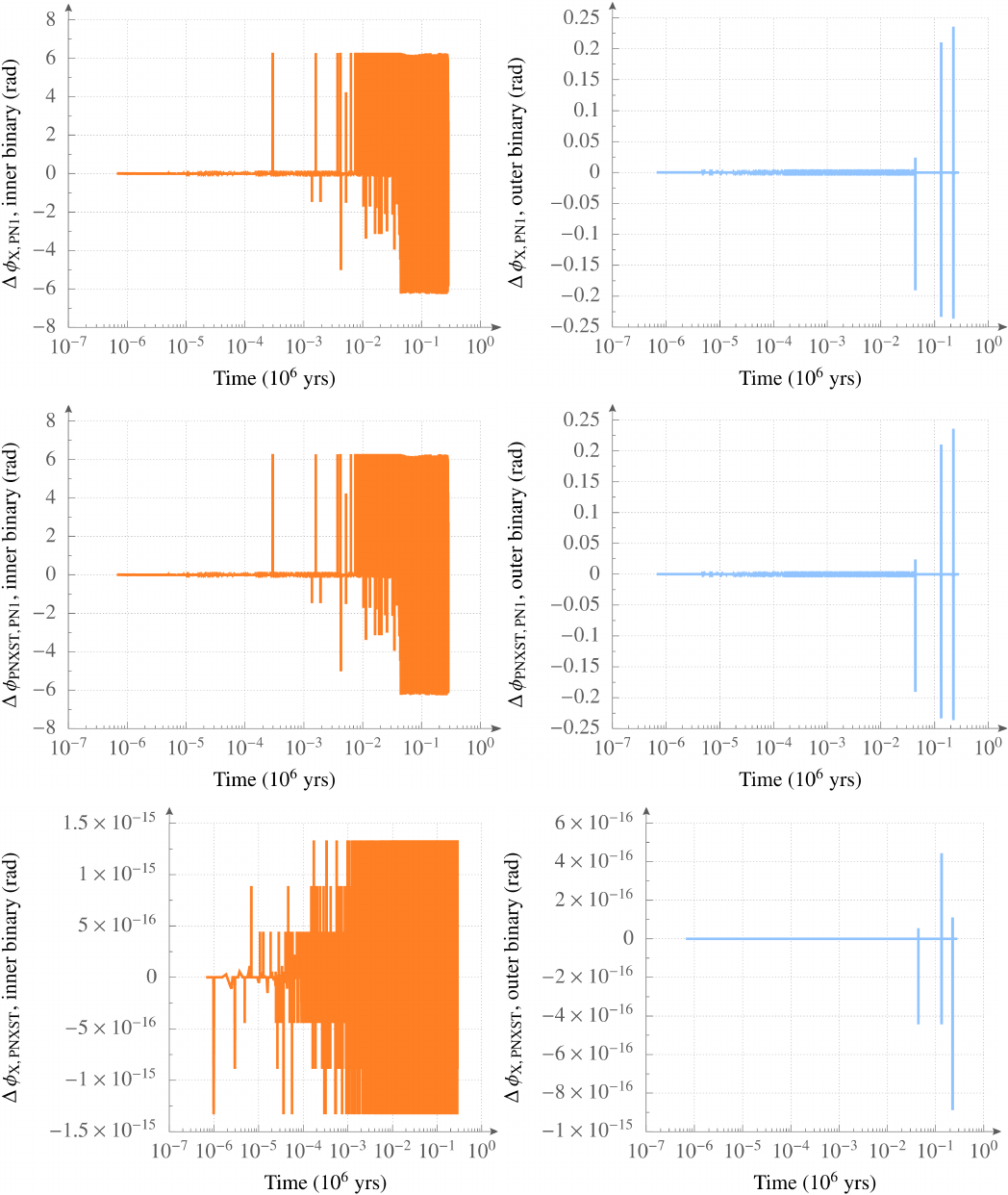}}
\caption
   {
Same as figure~(\ref{fig.Diff_Phase_E_EMRIs_Aproximats_low}) but for the system of an E-EMRI
and an X-MRI.
   }
\label{fig.Diff_Phase_E_EMRI_X_MRI_low}
\end{figure*}

In figure~(\ref{fig.Diff_Phase_E_EMRI_X_MRI_low}), we depict the same quantities as in figure~(\ref{fig.Diff_Phase_E_EMRIs_Aproximats_low}). We note that the phase shift for the outer binary tends to be smaller than in the previous case but still reaches relatively large values. Considering that the outer binary only completes three periapsis passages, this is not surprising and, at the same time, allows us to assess the impact of the cross terms, as even a small number of orbits leads to phase shifts of about $0.25$ rad, i.e., approximately 14$^\circ$. The inner binary reaches similar values as in the previous case.

\subsubsection{Duty cycle}

\begin{figure}
          {\includegraphics[width=0.5\textwidth,center]{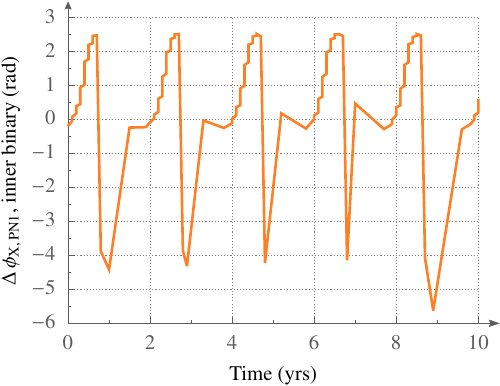}}
\caption
   {
Zoom in of the inner binary of Fig.~(\ref{fig.Diff_Phase_E_EMRI_X_MRI_low}) with the time reset to zero
to show the duty cycle of the system.
   }
\label{fig.phase_diff_inner_X_PN1_SHORT}
\end{figure}

It is particularly important to evaluate if the dephasing occurs in timescales shorter than
the observational time of LISA. We refer to this as the ``duty cycle'' of the effects coming
from the cross terms. We can readily see in Fig.~(\ref{fig.phase_diff_inner_X_PN1_SHORT}) a
close zoom-in of Fig.~(\ref{fig.Diff_Phase_E_EMRI_X_MRI_low}) at later times in the evolution
of the system which, as explained, represent a more realistic system. In the figure we have
reset the initial time to zero simply for clarity. We can readily see that the duty cycle
takes place in time periods of less than one year. This duty cycle is very close to the
rest of inner binaries we have addressed in this work.

\subsection{A system of two X-MRIs}

\begin{figure}
          {\includegraphics[width=0.5\textwidth,center]{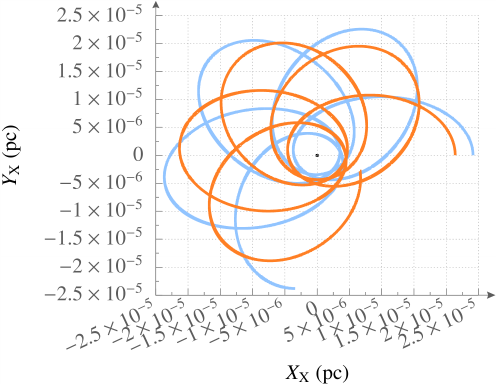}}
\caption
   {
The first few years in the evolution of two X-MRIs, each with a mass of \(0.02\,M_{\odot}\), orbiting an MBH of mass \(10^6\,M_{\odot}\) in the full cross-terms approximation. The initial semi-major axes are \(1.32\times 10^{-5}\) pc and \(1.42\times 10^{-5}\) pc, with eccentricities of \(0.62\) and \(0.70\), respectively.
}
\label{fig.OrbitsX_MRIs}
\end{figure}

In figure~(\ref{fig.OrbitsX_MRIs}), we show the early evolution on the XY plane of a system of two X-MRIs \citep{Amaro-Seoane2019} orbiting a supermassive black hole with the previously mentioned mass of $4.3\,M_{\odot}$. We do not display the full set of orbits, as it would appear as two concentric discs. We can readily follow the evolution in phase space of the two bodies revolving around the massive particle. We note that at the points of closest distance, such as the intersection at approximately $1.25\times 10^{-5}~\text{pc}$, the cross terms have the most significant effect. This is evident from Eq.~(\ref{acc1cross}), where the terms with the distance between the particles in the denominator dominate. However, since the initial dynamical configuration of the two particles is different, they do not necessarily cross that conjunction point at the same time.

\begin{figure*}
          {\includegraphics[width=1.0\textwidth,center]{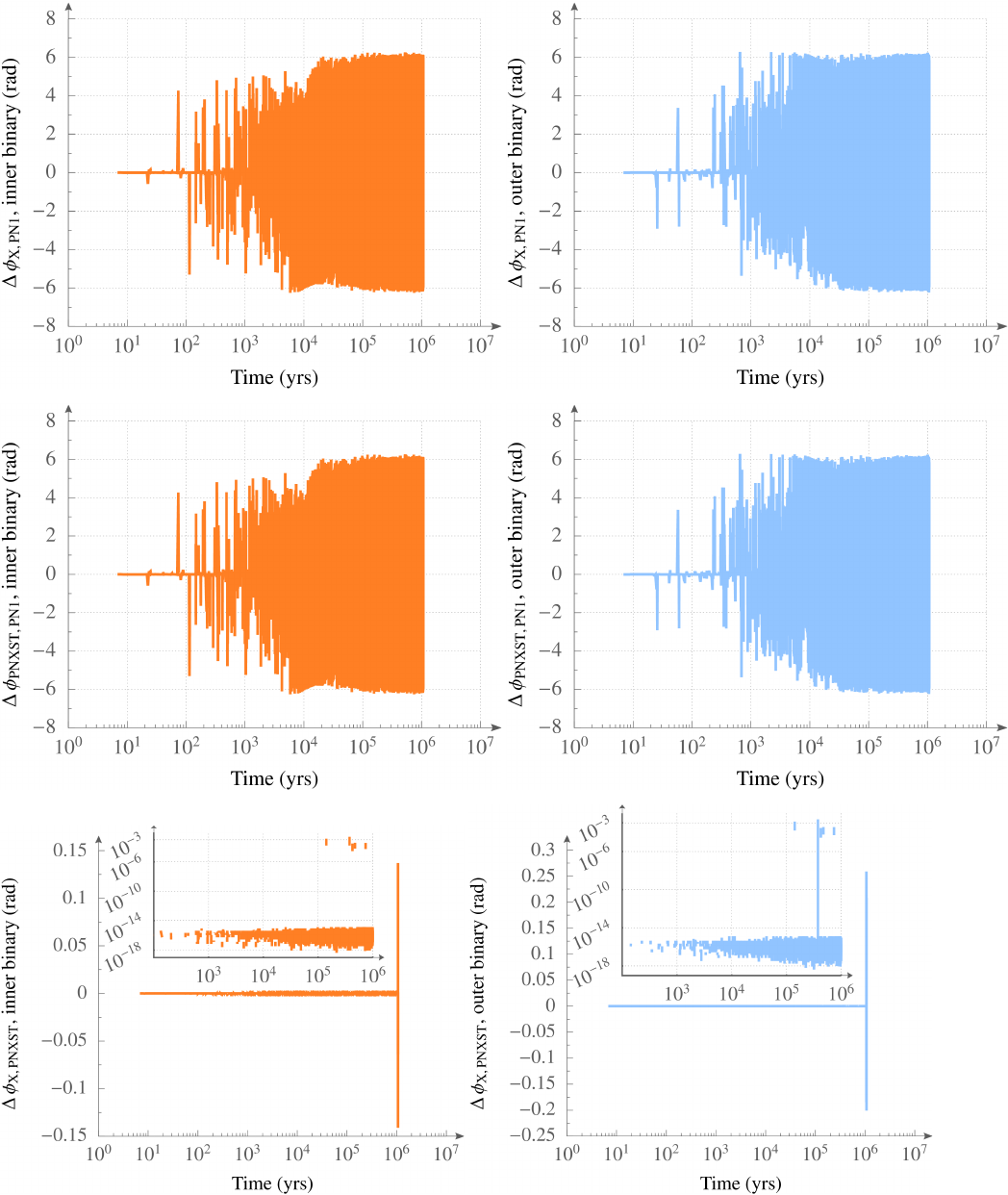}}
\caption
   {
Same as figure~(\ref{fig.Diff_Phase_E_EMRIs_Aproximats_low}) for the two X-MRIs configuration mentioned in figure~{\ref{fig.OrbitsX_MRIs}}. As usual, the difference between the standard PN1 treatment and the approach incorporating the cross terms results in phase shifts of approximately $2\pi$ (upper panels). Neglecting the correction for the MBH in the cross terms leads to a similarly significant phase shift (middle row). This difference is visible in the lower row of panels, particularly in the embedded magnified plots. Although the most significant spike occurs at later times, the binaries experience phase shifts of approximately $1.43^{\circ}$ when the cross terms between the stars and the MBH are neglected.
   }
\label{fig.Diff_Phase_XMRIs_low}
\end{figure*}

\section{Conclusions}

In this work, we have addressed the effect of the post-Newtonian cross terms, following the work of \cite{Will2014}, on perturbed asymmetric binaries with extreme- or extremely large mass ratios. Since the equations for the cross terms do not yet include dissipative terms and are limited to first-order corrections of the periapsis shift, we have investigated their impact on systems that shorten their semi-major axis very slowly, on timescales much longer than the observational time. From our point of view, these sources can be regarded as non-dissipative, ``monochromatic'' (in the sense that the peak of harmonics does not shift in frequency, although a cascade of harmonics accompanies the peak due to their initially large eccentricities). In that sense, we have studied systems consisting of two E-EMRIs orbiting a massive black hole \citep{AmaroSeoaneLinTzanavaris2024}, one E-EMRI with an X-MRI \citep{Amaro-Seoane2019}, or two X-MRIs. Although these cases are representative, they are merely illustrations of more complex situations. In particular, we have disregarded the fact that the massive black hole should host a swarm of approximately a few tens X-MRIs at any given time, introducing an additional layer of complexity to the system's dynamics. 

We have adapted a purely Newtonian, direct-summation $N-$body code to include both the first post-Newtonian correction term and the cross terms in the equations of motion for systems in which one body dominates in mass. The code we have chosen for this, \texttt{planet}, was developed for a similar situation, namely small bodies orbiting a star, with mass ratios comparable to those considered here. The programme exhibits good energy conservation with the implementation of the new equations of motion, typically reaching an error of $10^{-9}$ between timesteps in the corrector algorithm of the Hermite scheme. One might argue that the cumulative numerical error of an integration extending over millions of years would be problematic, but, as we explained in the relevant section, we are not comparing the final state with the initial conditions. 

We observe that the cross terms can lead to phase shifts of a few times $2\pi$, with typically a duty cycle of less than one year, which could render parameter extraction and possibly detection challenging. We have avoided episodic, transient, non-representative events that cause significant differences by removing all cases at $5\sigma$ from the distributions presented in the phase-shift results. Given the nature of the systems we have studied in our simulations, the solutions that should be closest to the steady-state case are those that have ``forgotten'' their initial conditions, i.e. the solutions at later times. When a space-borne observatory such as LISA begins to collect data, it is unlikely that these data originate from systems that formed only shortly before launch.

Our results suggest that any semi-Keplerian relativistic dynamical study involving more than two bodies should account for the cross terms; otherwise, the evolution will be incorrect. The initial scheme of \cite{KupiEtAl06}, which introduced post-Newtonian correction terms in the equations of motion for a system of $N$ bodies, although highly innovative, requires modification to produce accurate trajectories. Furthermore, there is a clear need for further investigation into the theoretical aspects of this problem, as higher-order (cross-term) corrections beyond the first term are still lacking.

\acknowledgments

We thank Cliff Will for discussions which have improved this work. We
acknowledge the financial support of the Generalitat Valenciana Project grant
CIAICO/2022/252. We dedicate this article to Sverre Aarseth, author of the
basic programme, \texttt{planet}. Sverre has been one of the cornerstones of
the numerical computation of stellar dynamics since its infancy. Apart from
being a guide in this respect, he was also a dear friend with whom many
intellectual and amusing moments were shared. Watching Sverre sculpt his
programmes live (because that's what he did, he chiselled the keyboard to near
destruction with the declaration of each variable) has been PAS's pleasure, and
a memory that has shaped him into the person he is.


\providecommand{\href}[2]{#2}\begingroup\raggedright\endgroup

\end{document}